\documentstyle[12pt]{article}
\begin{document}

\title{Shape transitions in proton-neutron systems}
\author{Serdar Kuyucak$^1$ \\
Department of Theoretical Physics,
Research School of Physical Sciences,\\
Australian National University, Canberra, ACT 0200, Australia \\[1cm]
Ka-Hae Kim$^2$ and Takaharu Otsuka$^3$ \\
Department of Physics, University of Tokyo, Hongo, Tokyo 113, Japan}
\date{}

\maketitle

\begin{abstract}
We explore the possibility of shape transitions in proton-neutron systems
driven by deformation differences between proton and neutron fluids. Within the
framework of the proton-neutron interacting boson model, we show that such
dynamic shape transitions cannot occur in well deformed nuclei but are a
possibility in transitional nuclei. Likely candidates in the Os-Pt isotopes are
discussed and predictions of the model are compared with the existing data. 
\\[.3cm]


\end{abstract}

\noindent
$^1$ E-mail: sek105@phys.anu.edu.au\\
$^2$ E-mail: kae@nt.phys.s.u-tokyo.ac.jp\\
$^3$ E-mail: otsuka@phys.s.u-tokyo.ac.jp
\vfill \eject

\baselineskip=21pt

Studies of shape transitions in nuclei at low excitation energies have been 
mostly based on coexistence of different shell model configurations 
\cite{woo92}.  In collective models, shape transitions could also arise 
from the competition between two macroscopic shape variables.  An example 
of such a shape transition, arising from the competition between the 
quadrupole and hexadecapole shapes, was given in the $sdg$ boson model 
earlier \cite{kuy91}.  Two-fluid models, where proton and neutron degrees 
of freedom are treated separately, could furnish another example.  
Measurements of neutron deformation using pion beams \cite{knu91} indicate 
that there could be sizable differences between the proton and neutron 
deformations.  There is also indirect evidence from g-factor measurements 
\cite{stu85} that the quadrupole operators for protons and neutrons could 
be rather different, especially in transitional nuclei \cite{kuy95}.  In 
this letter, we investigate in the framework of the proton-neutron 
interacting boson model (IBM-2) \cite{ots78,iac87} whether such differences 
could also drive a shape transition in proton-neutron systems.  For this 
purpose we use the analytic 1/$N$ expansion method \cite{kuy88} and the 
exact numerical diagonalization results from the code NPBOS \cite{ots85}.  
The 1/$N$ expansion is based on angular momentum projected mean field 
theory and therefore provides an intuitive picture for shape transitions.  
Further, due to their analytic formulation, a global study of the parameter 
space can be easily carried out, which shows that transitional nuclei 
provide the most likely candidates for such a dynamic shape transition.  As 
the 1/$N$ expansion results are not very reliable for transitional nuclei, 
we use exact diagonalization in exhibiting the shape transition and 
applications in this region.

We employ the simplest IBM-2 Hamiltonian suggested by microscopics
\cite{ots78,iac87}
\begin{equation}
H=\epsilon_\pi \hat n_{d\pi} + \epsilon_\nu \hat n_{d\nu}
- \kappa Q_\pi \cdot Q_\nu + \xi M,
\label{ham}
\end{equation}
where $\hat n_{d\rho}$, $\rho=\pi,\nu$ are the $d$-boson number operators
for proton and neutron bosons,
$M$ is the Majorana operator in Casimir form and
$Q_\rho$ are the quadrupole operators given by
\begin{equation}
Q_\rho= [d_\rho^\dagger s_\rho + s_\rho^\dagger \tilde d_\rho]
+\chi_\rho [d_\rho^\dagger \tilde d_\rho]^{(2)}, \quad \rho=\pi,\nu.
\label{q}
\end{equation}
The E2 matrix elements (m.e.) are calculated using the operator
\begin{equation}
T(E2)= e_\pi Q_\pi + e_\nu Q_\nu,
\end{equation}
where $e_\pi$, $e_\nu$ are effective charges, and the same quadrupole operators
(\ref{q}) are used as in the Hamiltonian.
For future reference, we introduce $F$-spin scalar and vector parameters as
\begin{eqnarray}
\epsilon_{\rm s}=(\epsilon_\pi+\epsilon_\nu)/2,\quad
\epsilon_{\rm v}=\epsilon_\pi-\epsilon_\nu, \nonumber\\
\chi_{\rm s}=(\chi_\pi+\chi_\nu)/2,\quad
\chi_{\rm v}=\chi_\pi-\chi_\nu.
\end{eqnarray}
Although other terms are sometimes included in detailed IBM-2 studies,
the Hamiltonian (\ref{ham}) is found to give an adequate description of
level energies and electromagnetic (E2 and M1) transitions \cite{iac87}.

The 1/$N$ expansion solutions follow from using the boson condensate
\begin{equation}
|N_\pi, N_\nu \rangle=(N_\pi! N_\nu!)^{-1/2}(b^\dagger_\pi)^{N_\pi}
(b^\dagger_\nu)^{N_\nu} |0\rangle, \quad
b^\dagger_\rho = (1+\beta_\rho^2)^{-1/2} (s_\rho^\dagger +
\beta_\rho d_{\rho0}^\dagger),
\end{equation}
as a trial state in a variation after projection (VAP) calculation.
Here $\beta_\pi$ and $\beta_\nu$ are the mean field deformations for the
proton and neutron fluids, which are determined from the expectation value
of the Hamiltonian (\ref{ham})
\begin{equation}
E_L= \langle N_\pi, N_\nu | HP^L_{00} | N_\pi, N_\nu \rangle/
\langle N_\pi, N_\nu | P^L_{00} | N_\pi, N_\nu \rangle
\label{el}
\end{equation}
by the VAP procedure.  In Eq.  (\ref{el}), $P^L_{00}$ denotes the angular 
momentum projection operator.  Note that the IBM deformation variables 
involve only the valence nucleons, and therefore they are much larger 
($\beta\sim 1-1.4$) than the typical geometrical model values ($\beta\sim 
0.2-0.3$) \cite{gin80}.  We refer to Ref.  \cite{kuy88} for details of the 
1/$N$ expansion method, and quote here the energy expression obtained from 
Eq.  (\ref{el}) for the Hamiltonian (\ref{ham}).  To the leading order in 
1/$N$, which is sufficient for discussion of systematic features, the 
energy surface and moment of inertia terms are given by
\begin{eqnarray}
&&\hskip -.5cm E_L= E_0 + C_L L(L+1), \\
&&\hskip -.5cm E_0= \sum_\rho {\varepsilon_\rho N_\rho \beta_\rho^2
\over 1+\beta_\rho^2}
-\kappa N_\pi N_\nu \beta_\pi \beta_\nu {(2+ \bar \chi_\pi \beta_\pi)
(2+ \bar \chi_\nu \beta_\nu) \over (1+ \beta_\pi^2)(1+ \beta_\nu^2)} +
{\xi N_\pi N_\nu (\beta_\pi-\beta_\nu)^2 \over
(1+ \beta_\pi^2)(1+ \beta_\nu^2)}, \\
&&\hskip -.5cm C_L={1 \over (aN)^2} \Biggl\{(6-a) \sum_\rho
{\varepsilon_\rho N_\rho \beta_\rho^2 \over 1+\beta_\rho^2} \\ \nonumber
&&\hskip .5cm + \kappa {N_\pi N_\nu \beta_\pi \beta_\nu \over
(1+ \beta_\pi^2)(1+ \beta_\nu^2)}
\Bigl(8a-12+(4a-12) (\bar\chi_\pi \beta_\pi+ \bar\chi_\nu \beta_\nu)
+(2a-9)  \bar\chi_\pi \bar\chi_\nu \beta_\pi \beta_\nu \Bigr) \\ \nonumber
&&\hskip .5cm +2\xi {N_\pi N_\nu (1+\beta_\pi \beta_\nu) \over
(1+ \beta_\pi^2)(1+ \beta_\nu^2)}
\Bigl( a+(a-6) \beta_\pi \beta_\nu\Bigr) \Biggr\},
\label{en}
\end{eqnarray}
where $\bar \chi_\rho=-\sqrt{2/7} \chi_\rho$. The quantity ``$a$'' represents
the average ``angular momentum squared'' carried by a single boson and is given
by
\begin{equation}
a={6 \over N} \left( {N_\pi \beta_\pi^2 \over 1+ \beta_\pi^2}
+ { N_\nu \beta_\nu^2 \over 1+ \beta_\nu^2} \right).
\end{equation}

In Fig.~1, we show the contour plots of the energy surface $E_0$ obtained 
from Eq.~(8) in a typical deformed (top) and transitional (bottom) nucleus.  
The positive $\beta$ values correspond to prolate and the negative ones to 
oblate deformations.  The prolate (absolute) minima are in the first 
quadrant and the oblate ones are in the third.  There are 20 contour lines 
at 1 MeV steps, thus the maxima in the second quadrant are about 20 MeV 
high.  In the deformed case, the prolate minimum in the first quadrant 
($E_{\min} = -4.83$ MeV) is well separated from the oblate one in the third 
quadrant with an energy difference of $\Delta E=2.23$ MeV. Including the 
moment of inertia term (9) in the energy surface, this energy difference 
gets larger with increasing spin.  Thus there is no chance of a cross over 
from the prolate to the oblate minimum.  Although we have used a particular 
parametrization here for illustration purposes, in fact, this is a general 
feature of all deformed nuclei as can be verified by systematic studies of 
Eqs.  (7-9).  Since such a study is already available for IBM-1 
\cite{kuy91}, and the IBM-2 results are very similar to those of IBM-1 for 
deformed nuclei, we will not elaborate on them further.

In the transitional case (bottom), the absolute minimum is at $E_{\min} = 
-3.07$ MeV, and the energy difference between the two minima is much 
smaller ($\Delta E=0.85 $ MeV).  Furthermore, the moment of inertia in the 
oblate minimum could be larger than that in the prolate one depending on 
the choice of the vector parameters in Eq. (4).  Hence the energy difference 
between the two minima could decrease with increasing spin, with 
possibility of a cross over at some critical spin.  Unfortunately, due to 
the soft energy surface, the 1/$N$ expansion formulas are not very reliable 
for transitional nuclei, and a quantitative study of shape transitions is 
not possible using the analytic formulas (7-9) in this region.

In order to exhibit the proposed proton-neutron shape transition and to 
study general conditions for its existence, we therefore rely on numerical 
diagonalization.  The signature for shape change is taken from the yrast 
spectroscopic quadrupole moments, which are negative for a prolate shape 
and positive for an oblate one.  Since the shape change is driven by 
differences in the proton and neutron deformations, the critical quantities 
to study are the $F$-spin vector parameters, $N_{\rm v}=N_{\pi}-N_{\nu}$, 
$\chi_{\rm v}$ and $\epsilon_{\rm v}$, which can induce asymmetries between 
the proton and neutron variables.  In principle, the Majorana interaction 
could also have an influence on the proton-neutron asymmetry.  However, 
variations in the Majorana strength (of order 20-30\%) did not lead to any
discernible effect on shape change.  In Fig.~2, we show the effect of these 
three vector parameters on yrast quadrupole moments in a typical prolate 
transitional nucleus with $F$-spin scalar parameters, $N=12$, 
$\epsilon_{\rm s}=0.4$ MeV, $\kappa=0.15$ MeV, $\xi=0.17$ MeV, and 
$\chi_{\rm s}=-0.2$.  The boson effective charges are taken as 
$e_\pi=e_\nu=0.1~eb$.  Fig.~2a shows the $N_{\rm v}$ dependence of the 
quadrupole moments $Q(L)$ for $\chi_{\rm v}=2$ and $\epsilon_{\rm v}=0$.  
It is seen that $Q(L)$ changes sign between the spins $L=16-18$ in the case 
of $N_{\rm v}=4$, but remains prolate in the other two cases.  Fig.~2b 
shows a similar study of $\chi_{\rm v}$ dependence of $Q(L)$ for $N_{\rm 
v}=4$ and $\epsilon_{\rm v}=0$.  Note that the bottom results in Figs.~2a 
and 2b are identical due to the symmetry of the Hamiltonian under the 
simultaneous interchange of $N_\pi-N_\nu$ and $\chi_\pi-\chi_\nu$.  Again 
only in the case of $\chi_{\rm v}=2$, a shape change occurs.  Finally, 
Fig.~2c shows the effect of $\epsilon_{\rm v}$ on the favourable case with 
$N_{\rm v}=4$ and $\chi_{\rm v}=2$.  Compared to $N_{\rm v}$ and $\chi_{\rm 
v}$, $\epsilon_{\rm v}$ plays a marginal role.  Nevertheless, it could 
shift the transition spin by several units in either direction depending on 
its sign; $\epsilon_{\rm v}<0$ makes the shape transition occur earlier 
while $\epsilon_{\rm v}>0$ retards it.  The opposite is true when $N_{\rm 
v}$ and $\chi_{\rm v}$ are both negative.

This systematic study indicates that an essential requirement for a dynamic
prolate-oblate shape transition to occur is the coherence of the two vector
parameters $N_{\rm v}$ and $\chi_{\rm v}$, that is they must have the same
sign. The $\epsilon_{\rm v}$ parameter with an opposite sign to the others
could also facilitate the shape transition though its effect is less important.
To understand these features better, we show in Fig.~3 the individual proton
and  neutron contributions to the yrast quadrupole moments in the favourable
($\chi_{\rm v}=2$) and the unfavourable ($\chi_{\rm v}=-2$) cases in Fig~2b.
In the favourable case (top), the neutron contribution changes little from
its original prolate value with increasing spin, while the proton
contribution sharply increases, changing sign at spin $L=12$.
At low-spins, the ($sd$) term in the quadrupole operator (2) dominates the
m.e. and it has the same  (negative) sign for protons and neutrons.
With increasing spin, however, the expectation value of the number of 
$d$-bosons in the states, and
hence the m.e. of the ($dd$) term in (2) increases while that of ($sd$) term
decreases. The sign of the quadrupole moment coming from the ($dd$) part is
determined by the sign of $\chi_\rho$ which is positive for protons and
negative for neutrons. Hence the ($sd$) and ($dd$) terms tend to cancel out
for protons but add up for neutrons, which explains the behaviour of the
individual proton and neutron quadrupole moments. The sign change in $Q(L)$
then follows from the fact that there are more protons than neutrons. In the
unfavourable case (bottom), $\chi_\pi$ and $\chi_\nu$ are interchanged, thus
protons favour the prolate shape and neutrons the oblate one in
high-spin states (the ground state remains prolate).
The behaviour of the individual contributions follow from the
same reasoning. Since there are more protons, they dominate the quadrupole
moment, and hence the nucleus stays in prolate shape.

The beneficial effect of $\epsilon_{\rm v}<0$ in hastening the shape 
transition can be understood in a similar manner.  It increases the 
deformation of protons and reduces that of neutrons, leading to a faster 
change in proton contribution to $Q(L)$ and reducing the contribution of 
neutrons.  In the more intuitive picture of Fig.~1 (bottom), including 
$\epsilon_{\rm v}=-0.4$ MeV in the Hamiltonian, reduces the energy 
difference $\Delta E$ between the two minima from 0.85 MeV to 0.68 MeV, 
hence making the cross-over from the prolate minimum to the oblate one 
easier.

These results for prolate transitional nuclei can be readily extended to
oblate ones by  noting that changing the sign of both $\chi_\pi$ and $\chi_\nu$
changes the sign of the quadrupole moments but their absolute values remain the
same. Thus in this case, dynamic oblate-prolate shape transitions would be
possible when $N_{\rm v}$ and $\chi_{\rm v}$ have the opposite signs. The
effect of $\epsilon_{\rm v}$ on $Q(L)$ remains similar; for $N_{\rm v}>0$,
$\epsilon_{\rm v}<0$ pushes the shape transition to an earlier spin, while
for $N_{\rm v}<0$, $\epsilon_{\rm v}>0$ does the same.

We use the above criteria in searching for a potential candidate among the 
transitional nuclei that might exhibit such a dynamic shape transition.  
The $F$-spin vector parameters in Os-Pt isotopes were recently determined 
from a study of $M1$ properties, resolving the long standing anomalies 
observed in the $g$-factors of these nuclei \cite{kuy95}.  Since $N_{\rm 
v}<0$ in all the isotopes considered in Ref.  \cite{kuy95}, a negative 
value of $\chi_{\rm v}$ in Os isotopes and a positive one in Pt isotopes is 
a necessary condition for a dynamic shape transition.  In addition, a 
positive value of $\epsilon_{\rm v}$ would help it to occur at an earlier 
spin.  Inspection of Table 1 in Ref.  \cite{kuy95} shows that $^{192}$Os 
provides the most favourable case for a prolate-oblate shape transition.  
In Fig.~4, we show the level energies, E2 transition m.e.  and quadrupole 
moments in the ground band of $^{192}$Os, obtained using the same 
parameters as in Ref.  \cite{kuy95}.  Although the calculated quadrupole 
moments do not actually change sign, they depict the same rapid change seen 
in Fig.~3, and come very close to doing so.  The failure is due to the 
relatively low boson number ($N=8$), which causes the boson cutoff effect 
to kick in early.  The data on E2 transitions (Fig.~4) follow the axial 
rotor results closely, and do not show any sign of boson cutoff.  Thus the 
premature decrease in the E2 m.e.  is really a problem of the $sd$-IBM, and 
for a proper description of high-spin states one needs to include 
$g$-bosons in the basis.  Unfortunately, an exact $sdg$-IBM-2 calculation 
for this nucleus is not feasible at present due to the large basis space.  
An approximate calculation in a truncated basis space is possible but, as a 
recent study indicates \cite{li96}, truncation leads to unreliable results 
for spins $L>2N$.  This is exactly where the $sdg$-IBM-2 results are 
needed, therefore truncated calculations will not serve a useful purpose 
for extending the present results to higher spins.  Nevertheless, both the 
proton-neutron deformation difference and the hexadecapole effects drive 
the system in the same direction (from prolate to oblate shape), and 
inclusion of g bosons could only enhance the shape transition.  Thus 
combination of the present $sd$-IBM-2 results with the earlier $sdg$-IBM-1 
ones \cite{kuy91,lac92}, strongly suggests a shape transition in the spin 
range $L=10$ to 20 for $^{192}$Os.  The measured quadrupole moments deviate 
markedly from the axial rotor results, but to ascertain whether they 
actually decrease in absolute value would require more accurate 
measurements.

In the absence of any other guiding principle, we used the vector 
parameters that were determined from the $M1$ properties in our search for 
shape transitions.  While the $F$-spin breaking mechanism in the IBM-2 
appears to give a mostly consistent description of $M1$ properties in 
rare-earth nuclei, it is also argued that $M1$ data are really sensitive to 
single particle degrees of freedom and should not be used to determine 
collective variables \cite{lev90}.  From this point of view, observation of 
dynamic shape transitions (or lack of it) would help to determine the 
vector parameters in an independent way, and hence it would provide a 
significant test for the collective model interpretation of the $M1$ 
observables used in the IBM-2.  Structurally, the quadrupole parameters 
used for $^{192}$Os ($\chi_\pi=-0.68$, $\chi_\nu=0.32$) imply competing 
prolate and oblate shapes for the proton and neutron fluids, respectively.  
This picture is similar to the pairing-plus-quadrupole model calculations 
\cite{kum72}, which first predicted the static prolate-oblate shape-phase 
transition in Os-Pt nuclei.

The errors in the measured quadrupole moments of Os-Pt isotopes \cite{wu96}
are generally too large to reach a definite conclusion whether they actually
decrease in absolute value with increasing spin.
More precise measurements of the quadrupole moments extending to higher spins
are needed to test this  prediction. Such measurements are now possible with
the new 4$\pi$ high-resolution detector systems ``Euroball" and
``Gamma-Sphere". It would be very interesting to see whether collective nuclei
actually exhibit dynamic shape transitions driven by differences between
macroscopic shape variables. Apart from that, they would also provide an
important consistency check on the $F$-spin breaking mechanism used in
explaining the $M1$ properties in the IBM-2.

S.K. thanks the members of the nuclear theory group at the University of Tokyo,
where this research was carried out, for their hospitality.
This work was supported in part by the Australian Research Council and the
Department of Industry, Science and Technology, and in part by the Japanese
Ministry of Education, Science and Culture.

\eject

\vfill \eject
{\Large \bf Figure captions}
\\[.5cm]
Fig. 1. Contour plots of the energy surface (8) in the $\beta_\pi-\beta_\nu$
plane for a typical deformed nucleus (top) and transitional nucleus (bottom)
with $N_\pi=8$, $N_\nu=4$ bosons.
The parameters are (in MeV except for $\chi$) $\varepsilon_{\rm v}=0$,
$\kappa=0.15$, $\xi=0.17$, $\chi_{\rm v}=2$ for both, and
$\varepsilon_{\rm s}=0.2$, $\chi_{\rm s}=-0.5$ in the deformed and
$\varepsilon_{\rm s}=0.4$, $\chi_{\rm s}=-0.2$ in the transitional case.
The prolate (absolute) minima are in the first quadrant and the oblate ones
in the third. The contour lines are separated by 1 MeV, and the absolute
minima are $E_{\min} = -4.83$ MeV (top), $E_{\min} = -3.07$ MeV (bottom).
\\[.4cm]
Fig. 2. Systematic study of the yrast quadrupole moments in a transitional
nucleus against the $F$-spin vector parameters; $N_{\rm v}$ (a),
$\chi_{\rm v}$ (b), and $\varepsilon_{\rm v}$ (c). The scalar parameters are as
in Fig.~1 (bottom). The vector parameters are $\chi_{\rm v}=2$,
$\varepsilon_{\rm v}=0$ (a), $N_{\rm v}=4$, $\varepsilon_{\rm v}=0$ (b), and
$N_{\rm v}=4$, $\chi_{\rm v}=2$ (c). The lines are drawn to guide the eye.
\\[.4cm]
Fig. 3. Proton-neutron decomposition of the yrast quadrupole moments
for the two cases with $\chi_{\rm v}=\pm 2$ in Fig.~2b. The other vector
parameters are $N_{\rm v}=4$ and $\epsilon_{\rm v}=0$.
The proton and neutron contributions are indicated by open circles, and the
total quadrupole moment by filled circles. The lines are drawn to guide the eye.
\\[.4cm]
Fig. 4. Comparison of the energies, E2 transitions and quadrupole moments
of the ground band states in $^{192}$Os calculated in IBM-2
(solid lines) with the experimental data (circles) \cite{wu96}.
The parameters (taken from Ref. [5]) are; $N_\pi=3$, $N_\nu=5$,
$\varepsilon_{\rm s}=0.4$, $\varepsilon_{\rm v}=0$, $\kappa=0.15$,
$\chi_{\rm s}=-0.18$, $\chi_{\rm v}=-1$, $\xi=0.17$ (in MeV except for
$N$ and $\chi$), and $e_\pi=e_\nu=0.15~eb$. The results for an axial rotor with
$Q_0=4.6~eb$ are also shown for comparison (dashed line).

\end{document}